\documentclass{llncs}
\usepackage{amscd}
\usepackage{amsfonts}
\usepackage{amssymb}
\usepackage{graphicx}
\usepackage{amsmath}
\begin{document}
\title{A Meaning-oriented Approach to \\Semantic Data Modeling}
\author{Xuhui Li\inst{1}}
\institute{School of Information Management, Wuhan Univ.,\\
Wuhan, China\\ \email{lixuhui@whu.edu.cn}  
} \maketitle

\begin{abstract}

Semantic information is often represented as the entities and the relationships among them with conventional semantic models. This approach is straightforward but is not suitable for many posteriori requests in semantic data modeling. In this paper, we propose a meaning-oriented approach to modeling semantic data and establish a graph-based semantic data model. In this approach we use the meanings, i.e., the subjective views of the entities and relationships, to describe the semantic information, and use the semantic graphs containing the meaning nodes and the meta-meaning relations to specify the taxonomy  and the compound construction of the semantic concepts. We demonstrate how this meaning-oriented approach can address many important semantic representation issues, including dynamic specialization and natural join.

\end{abstract}

\section{Introduction}

Semantic information management is an interesting topic in the fields of data management and knowledge engineering. To effectively manage and use the semantic information, an expressive semantic data model is required as the theoretical foundation, thus it has been a central research topic since the late 1970s. Researchers have proposed various semantic data  models generally following the entity-relationship-oriented approach. The essence of this approach is to describe the semantics of the entities and the relationships among them in a properly organized structure, since the semantic information is always about the entities and their relationships.

Under the guidance of the entity-relationship-oriented approach, the existing studies put the emphasis on different aspects of the entity-relationship couple. Some literatures prefer to model the entities (or the ``objects'' in O-O modeling), and use the ``attributes'', the meaningful functions between the entities, to indicate the semantic information. Some literatures are apt to model the relationships, and use the type constructors, especially the the ``aggregation'' representing the n-ary relationships among the entities, to do it. Also, there are some studies on the graph data models which use the nodes and edges to represent the entities and relationships respectively and show special interests in the connectivity information.

The entity-relationship-oriented approach often motivates the designers to think about the entities and relationships in an ontological way, that is, what an entity is and how a relationship associates the entities. As the consequence, the above data models often provide the rigid taxonomy mechanisms which require the users to have a considerate and ``objective'' view to the semantic features to model, as the common term ``object'' indicates. These models can work well in a priori assumption, i.e., the application requests are known, so that the entities and the relationships are modeled according to the considerate view of the application requests. However, the modeling results are essentially the subjective views of the designers to the things in application. For example, the attributes of an entity are gathered in a pragmatistic way according to the requests. When the application requests are changed and are not satisfied by the predefined schemas, the models can hardly provide a flexible and expressive mechanism to modify and reuse the schemas to satisfy these ``posteriori'' requests.

In this paper, we propose a meaning-oriented approach to semantic modeling to overcome the above insufficiencies. In this approach, we  use the meanings, i.e., the subjective views of the things  including the entities, the relationships and the activities, etc., as the basic units in modeling semantic information.
The major advantage is that, since the meanings are essentially subjective, designers can focus on their own consideration of the things to model and can utilize the modeling mechanisms to reuse, extend and integrate the meanings for new application requests. Our contributions in this paper include: 1) We propose a graph-based data model named SemGraph following the meaning-oriented approach. This ongoing study adopts a simple but expressive graph structure with the fundamental meta-semantic relations and the corresponding mechanisms to present the complex features of semantic information. 2) We explore the typical semantic modeling methods based on the new data model, and the results show that our model can undertake all the common semantic construction tasks in a more flexible way.


The remainder of this paper is organized as follows.  Section 2  briefly introduces the related work of the data models involved in semantic modeling. Section 3 describes the data structure and the schema mechanisms of the SemGraph, which lays the foundation of the semantic data modeling using the meaning-oriented approach. Section 4 discuss the major features of SemGraph in various important issues of semantic modeling. Section 5 concludes the paper.

\section{Related Work}

As mentioned previously, the early semantic models often follow two philosophical approaches\cite{survey1,survey2}: one approach places an emphasis on explicitly type constructors\cite{ifo}, and the other stresses the use of attributes to interrelate objects\cite{sdm}.
The later studies on data modeling were overwhelmingly affected by the rapid development of object-oriented paradigms, which mainly follows the attribute-based approach. Various object-oriented data models \cite{3,4,16} have been proposed to model the semantic objects.  Generally, these models often adopt taxonomical approach to classify the semantic information into various categories corresponding to various concepts, and use various semantic features such as object identity, aggregation, classification, instantiation, generalization/specialization, class hierarchies, non-monotonic inheritance,etc., to describe and refine the concepts. They are mainly concerned about the static aspects of the things and normally require an object to be an instance of a most specific class.

However, the semantic information in the real world is often dynamic and interrelated, which makes the semantics evolution a common case. To deal with the dynamic and many-faceted aspects of real-world objects, various role models
 have been proposed \cite{13}. The main characteristics of
these role models is the separation of object classes and role classes so that
an object can play several roles. Further, the context of the roles is also important in semantic modeling\cite{inm}.

The interrelationship among semantic objects was firstly modeled in semantic networks and has been studied in various graph data models\cite{gsurvey1}. Some of them concern the semantic information embodied in the nodes, the edges and even subgraphs or hypernodes\cite{gdm,groovy}. These models inherited more or less features from conventional semantic data models and object-oriented models, and use them to specify the concepts in a graph-oriented way. Even though, these models are seldom suitable for some important issues of practical semantic modeling, e.g., the posteriori semantic evolution and the dynamic roles.
Another kind of graph data models aim at establishing and querying a graph database with specific semantic information. Some of them can present simple semantic features like aggregation\cite{gdm1} and attribute\cite{rdfs}, e.g., the attribute-oriented RDFS model for building the Semantic Web\cite{lod}, or graph-oriented features like link and path\cite{graphdb}. The others often adopt very simple graph structure, i.e., nodes and edges containing certain values, and focus on the efficiency of processing typical graph queries in certain kinds of graph datasets\cite{gsurvey2,graphdata}.

The recent decade has also seen the popularity of document databases, i.e., the data sets of XML or JSON-like documents. These databases often contain and manipulate the semi-structured semantic information. For example, the tags or the keys of the data elements are often regarded as semantic labels of the data, and the hierarchically organized data schema can also present the compound structure of the semantic concepts. However, studies often concern using these documents to implement the common semantic models like OWL or the specific models\cite{ghafoor}, and only some preliminary work \cite{xmlsem}  has been done on building semantics models on these hierarchically-organized  documents.

\section{SemGraph: A Meaning-oriented Semantic Data Model}
\subsection{The philosophy and the data structure}

SemGraph adopts an idealistic philosophy to specify the semantic information. We find that, when we are talking about the concrete things in the real or conceptual world, what really affects us is just the subjective impression. An example is that, for a novel where the events and people are all fictional, there is often a sound semantic world  imagined by the readers. Therefore,
if we need a proper semantic model to describe the world, it should be enough to build a sound and expressive system for representing our subjective impression and our language. This viewpoint has already been verified by the ``epistemological turn'' and the ``linguistic turn'' in philosophy.

Following this philosophy, the semantic world is completely built on the subjective impression. Therefore, we neither directly describe the objective entities and the relationships nor differentiate the entity and the relationships. Instead, we use the ``meaning instance'' to denote the subjective impression in one's mind when he thinks about a thing, e.g., an entity, a relationship or an activity,  in the real or imaginary world. For example, to describe a school entity, we don't consider what the school is, but dwell on the information about the school from a certain viewpoint, e.g., teaching or geo-information. Consequently, the semantic information is treated as the interrelated meaning instances. To exchange and understand the subjective impressions from different minds, the meaning instances have to be categorized into certain taxonomical types, i.e., subjective views or concepts; and to organize and manage these meaning instances, the interrelations among them  also need to be categorized into certain meta-semantic relations and be represented structurally.

SemGraph is a semantic data model following the meaning-oriented approach. It adopts a directed graph structure, called \emph{meaning graph}, to represent the semantic information. Different from many graph data models using the nodes and the edges to respectively represent the entities and the relationships, SemGraph only uses the labelled nodes to store normal semantic information, i.e., the meaning instances. The label of a node indicates its meaning type, and it is also called the ``meaning'' in short. Therefore, a node is specified in the form of ``\emph{id:Label}'' where the ``\emph{id}'' denotes the instance and the ``\emph{Label}'' denotes the meaning. Especially, for a value meaning label like ``\emph{\#String}'', the \emph{id} of a meaning node is the value of the meaning instance, e.g.,  ``\emph{`Joe':\#String}''.
SemGraph uses the labelled edges to represent the meta-semantic relations. In SemGraph we mainly concern four kinds of meta-semantic relations: the ``\emph{composition}'', the ``\emph{equivalence}'', the ``\emph{specialization}'' and the ``\emph{reference}'' relations. These relations are respectively represented by the meta-semantic labels ``$\rightarrow$'', ``$\Leftrightarrow$'', ``$\Leftarrow$'' and``$\dashrightarrow$'' of the edges.
A meaning graph can represent meaningful semantic information only if its meta-meaning relations are reasonably and consistently designed and the graph can satisfy the structural and logical constraints imposed to the meta-meanings.
How to specify and use the meta-semantic relations to present the complex semantic features is thus the central issue of the SemGraph modeling, as we would show in the rest of the paper.

\subsection{Composition and equivalence}
Fig.1 illustrates the running example, a meaning graph of a school, from the ``teaching'' viewpoint. This graph indicates that the teachers like the person ``\emph{Joe}'' and the students like the person ``\emph{John}'' in the school be the certain specializations of the teachers and the students in a general teaching relationship, meanwhile the school administrates its own group teaching relationships where the students' study is assessed with the scores or grades. For brevity, it uses the abbreviated meaning labels, e.g, ``\emph{Teacher}'' or ``\emph{Student}'', in the nodes.

\begin{figure}[h]
\begin{center}
\includegraphics[width=.8\textwidth]{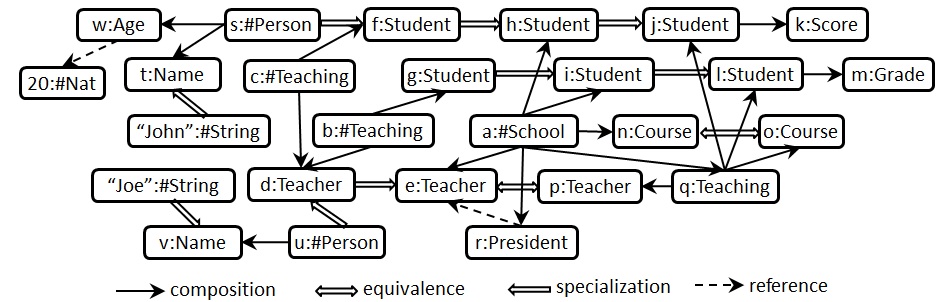}
\caption{An example of meaning graph}
\end{center}
\end{figure}

As shown in Fig.1, the most fundamental meta-meaning relation is the composition relation presenting the composite structure of a meaning.
In a composition relation  ``\emph{a:\#School$\rightarrow$e:Teacher}'' in Fig.1, the meaning ``\emph{\#School}'' of the source node \emph{a} is a composite meaning and the meaning of the target node \emph{e} with the abbreviated label  ``\emph{Teacher}'' is its component (meaning).

A component of a composite meaning can itself be a composite meaning, called a composite component, thus forming a composition hierarchy. A composition hierarchy is a directed acyclic graph of the meanings where the edges represent the composition relations. A meaning being not the component of the other meaning is called a root meaning.

SemGraph adopts a naming convention to structurally specify the meaning labels and the composition hierarchy. A meaning name is a string in the style of ``\emph{\#root.component}'' which starts with a ``\#'' symbol denoting the root meaning name and uses the ``.'' symbol as the infix to indicate the composition. The abbreviated label like ``\emph{component}'' or ``\emph{.component}'' is used to indicate that the nodes it denotes be under the context of the current composite node or root node. For example, the meaning name (or label) ``\emph{\#School.Teacher}'' can be abbreviated as  ``\emph{Teacher}'' under the context of ``\emph{\#School}'' as Fig.1 shows, and it can also be abbreviated as ``\emph{.Teacher}'' under the context of ``\emph{\#School.Teaching}''.


The components of a composite meaning should follow a certain composition schema. In a composition schema, the component labels are declared with their own composition schemas, and each component label \emph{l$_{c}$} is attached with a scope modifier and an optional set-of mark ``[*]''. Furthermore, the component labels are presented in an occurrence expression indicating the set of the valid occurrence combination of the component labels. For example, the composition schemas of the meanings ``\emph{\#Teaching}'' and ``\emph{\#School}'' in Fig.1 is listed as below:

\begin{small}\begin{verbatim}
meaning #Teaching -> {public Teacher, public Student}
meaning #School -> {
    Teacher[*], Student[*], Course[*], President,
    Teaching[*] -> {
        public Teacher <=> .Teacher,
        public Course <=> .Course,
        Student[*] -> {(Grade | Score)?} } ...  } ...
\end{verbatim}\end{small}

A component meaning \emph{l$_{c}$} can be modified as ``\emph{static}'', ``\emph{public}'' or ``\emph{private}'' (on default): a static component \emph{l$_{c}$} is shared, i.e., being pointed to with the ``$\rightarrow$'' edges from multiple composite nodes, by all the nodes of its composite meaning \emph{l}; a public component \emph{l$_{c}$} is allowed to be shared by multiple nodes of \emph{l}; and a private component  \emph{l$_{c}$} only pertains to one node of \emph{l}. As shown in the example, the meaning ``\emph{\#Teaching}'' has two public components ``\emph{Teacher}'' and ``\emph{Student}''. That means, the ``\emph{\#Teaching.Teacher}'' and the ``\emph{\#Teaching.Student}'' nodes can be shared by multiple ``\emph{\#Teaching}'' nodes.
The meaning ``\emph{\#School}'' has the components ``\emph{Teacher}'', ``\emph{Student}'', ``\emph{Course}'', ``\emph{President}'' and ``\emph{Teaching}''. All the components are private, i.e., they only make sense in their own ``\emph{school}''. They are declared as the set-of components except for the ``\emph{President}'', indicating that there be one or more component nodes with the same name.

The component labels are usually embedded in an occurrence expression in the composition schema. The abstract syntax of the occurrence expression \emph{oe} is\\
\indent \indent \indent\indent \emph{oe ::=  l $|$ `('oe`)' $|$ oe`?' $|$ oe`,'oe $|$ oe`$|$'oe}\\
where \emph{l} deonte the meaning labels.
Each component label can occur only once in the occurrence expression. As its syntax shows, there are three kinds of occurrence for the component labels, i.e., conjunctive, disjunctive, or optional, respectively denoted using the operator ``,'', ``$|$'' and ``?''. For the conjunctively combined labels, the components should occur simultaneously under the same context; for the disjunctively combined labels, the components  occur mutually exclusively; for the optional labels, the components may or may not occur. An occurrence expression actually defines an occurrence set \emph{O$_{l}$} whose elements are the sets of the component labels. Such a label set indicates a valid co-occurrence of the components. For example, in the meaning ``\emph{\#Teaching}'' and ``\emph{\#School}'', their occurrence sets only contain one element which includes all the components; in the meaning ``\emph{\#School.Teaching.Student}'', the component ``\emph{Score}'' and the component ``\emph{Grade}'' are both optional and they should not occur together, and thus the occurrence set is ``{\{\emph{\{\}, \{Grade\}, \{Score\}}\}}''.

As shown in Fig.1, there is an equivalence relation  ``\emph{e:\#School.Teacher $\Leftrightarrow$ p:\#School.Teaching.Teacher}''. The equivalence relation indicates the two meaning nodes are extensionally equivalent. It is an isomorphism between the meaning graphs whose meaning labels may or may not be the same.

SemGraph currently deploys a conservative mechanism to specify the equivalence relation in two aspects. On one hand, the equivalence between meanings can be specified using the equivalence declarations in the form of ``\emph{l$_{1}$ $\Leftrightarrow$ l$_{2}$}'', e.g., the declaration ``\emph{Teacher$\Leftrightarrow$.Teacher}'' in the schema of ``\emph{\#School.Teaching}''. For \emph{l$_{1}$} and \emph{l$_{2}$}, their components should also be equivalent accordingly, and thus the declarations are often organized hierarchically. In practice, the two equivalent nodes can often be combined if their components all share the same local labels.

On the other hand, for the equivalence between the meaning nodes like \emph{n:l} and \emph{n':l},  at current stage a rigid mechanism is adopted to declare the meaning \emph{l} as ``\emph{intensive}'', indicating the equivalence between the two nodes holds if their components are accordingly in equivalence relations. This mechanism is useful to define the value meaning types, e.g., ``\emph{\#Nat}'' or ``\emph{\#String}'', as shown later, and a more powerful equivalence mechanism is being explored.

\subsection{Specialization and reference}

The specialization relation between two meaning nodes indicates that
the target meaning instance be a specialization of the source meaning instance.  In  the relation ``\emph{e:\#School.Teacher$\Leftarrow$d:\#Teaching.Teacher}'' in Fig.1, the meaning ``\emph{\#Teaching.Teacher}''  is a \emph{super meaning} of the meaning ``\emph{\#School.Teacher}'', and the label ``\emph{\#School.Teacher}'' is also named a \emph{sub meaning} of the meaning ``\emph{\#Teaching.Teacher}''. The source node and the target node are thus respectively called the \emph{super-node} and the \emph{sub-node}.

The sub meaning should inherit the component meanings from the super meaning meanwhile it can extend the super meaning by introducing new component meanings. The inherited components are essentially the equivalent meanings of the ones in the super meaning. The sub meaning can also explicitly use the specialization of the inherited component meanings, which can even be with the same names to override the inherited ones. For example, for the composite meaning ``\emph{\#School}'' and its sub meaning ``\emph{\#University}'', the component meaning ``\emph{\#School.Teacher}'' is inherited as ``\emph{\#University.Teacher}'' and can be specialized as the new component meanings ``\emph{\#University.Professor}'' or ``\emph{\#University.Lecturer}''. Furthermore, the new component meaning ``\emph{\#University.Teacher}'' can be defined to override the ``\emph{\#School.Teacher}''.

SemGraph uses the specialization declaration in the form of ``\emph{l$_{1}$$\Lleftarrow$l$_{2}$}'' to explicitly specify the specialization relation between the meanings. The notation ``$\Lleftarrow$'' generalizes the three kinds of specialization styles,  the normal, the sub-node-restriction and the super-node-restriction styles respectively denoted by the symbols ``$\Leftarrow$'', ``$\Leftarrow$!'' and ``!$\Leftarrow$''. For example, the specialization declarations in the ``\emph{\#School}'' meaning is listed as below
\begin{small}\begin{verbatim}
meaning #School -> { ...
    Teaching[*] -> {...} with { Student <= .Student; ...} }
with { Teacher !<= #Teaching.Teacher;
       Student !<= #Teaching.Student; ...}
\end{verbatim}\end{small}

For a normal specialization \emph{l$_{1}$$\Leftarrow$l$_{2}$}, the node of the super meaning \emph{l$_{2}$} and the node of the sub meaning \emph{l$_{1}$} are relatively independent to each other, because a sub-node just represents a subjective extension to the super-node  and thus the existence of the sub-node should not affect the super-node itself. Therefore, a super-node can have more than one sub-nodes, even the sub-nodes are of the same meaning types, meanwhile a sub-node can also have different super-nodes, while the super-nodes should have different types unless they are in the equivalence relation. As shown in the above declarations, the nodes of ``\emph{\#School.Teaching.Student}'' is relatively independent to their super-nodes of the ``\emph{\#School.Student}'' meaning.

%

For a sub-node-restriction specialization \emph{l$_{1}$!$\Leftarrow$l$_{2}$}, a node of the super meaning \emph{l$_{2}$} can only have one node of the sub meaning \emph{l$_{1}$}. For a super-node-restriction \emph{l$_{1}$$\Leftarrow$!l$_{2}$}, a node of \emph{l$_{2}$} can only have one node of l$_{1}$ and further, the super-node exists only if the sub-node exists. That is, the super node would be deleted if the sub node is deleted, and so do the other sub nodes.
These two restrictions are used to emulate the object instance mechanisms in conventional O-O modeling where the instances of the super-type and the sub-type are combined as one instance. As the above example shows, a node of the ``\emph{\#School.Teacher}'' meaning should exist with its super-node of ``\emph{\#Teaching.Teacher}''.

For the specialization relation \emph{a:l$_{1}$$\Leftarrow$b:l$_{2}$}, since the component meanings of \emph{l$_{1}$} and \emph{l$_{2}$} are in equivalence or specialization relations accordingly, there should be a surjective map from the component nodes of \emph{a} to the ones of \emph{b}. Therefore, it is required that the composition relations should maintain in the specialization, especially, the component occurrence set of the sub meaning should be consistent with the component occurrence set of the super meaning.
We say a sub meaning \emph{l$_{1}$} \emph{is consistent with} its super meaning \emph{l$_{2}$} with respect to the component occurrence, if for each valid label occurrence \emph{o'} in the occurrence set of \emph{l$_{1}$}, the super meanings of the components in \emph{o'} can form a valid label occurrence \emph{o} in the occurrence set of \emph{l$_{2}$}. SemGraph has the following property to check the validity of a graph specification with respect to the component occurrence consistency:

%

\begin{proposition}
For the two meanings l$_{1}$$\Lleftarrow$l$_{2}$ with their composition schemas and the specialization declarations, whether the specialization between l$_{1}$ and l$_{2}$ is consistent with respect to the component occurrence is decidable in PTIME.
\end{proposition}

A composite meaning often contains the component meanings which are inherited not from its super meanings but from the component meanings of other meanings. Such a situation is called the component specialization. For example, the meanings ``\emph{\#School.Teacher}'' and ``\emph{\#School.Student}'' are the sub meanings of ``\emph{\#Teaching.Teacher}'' and  ``\emph{\#Teaching.Student}'' respectively.  For a ``\emph{\#School}'' meaning node  containing the ``\emph{\#School.Teacher}'' and the ``\emph{\#School. Student}'' component nodes, there should be certain ``\emph{\#Teaching}'' nodes, like \emph{b} and \emph{c} in Fig.1,  whose component nodes are the super-nodes of the ``\emph{\#School.Teacher}'' and ``\emph{\#School.Student}'' nodes.

The above example shows a possible completeness problem incurred by the the component specialization. That is, how are the implicit ``\emph{\#Teaching}'' meaning nodes constituted. This problem is  addressed by a special constraint named the complement constraint. This constraint uses two kinds of graph pattern expressions, the component graph patterns and the component-specialization graph patterns.
 A component graph pattern is an expression \emph{p$_{c}$} defined as \\
\indent\indent\indent\indent \emph{p$_{c}$ ::= v:l $|$ (p$_{c}$) $|$ p$_{c}$,p$_{c}$ $|$ v:l $\rightarrow$ p$_{c}$},\\
 and a component-specialization graph pattern is an expression \emph{p$_{cs}$} defined as \\
\indent\indent\indent\indent \emph{p$_{cs}$ ::= r $|$ (p$_{cs}$) $|$ p$_{cs}$,p$_{cs}$ $|$ r $\rightarrow$ p$_{cs}$},\\
\indent\indent\indent\indent \emph{r ::= v:l $|$ v$\Leftarrow$v:l $|$ v:l$\Leftarrow$v}\\
where \emph{l} denotes the meaning labels and \emph{v} denotes the variables.

A complement constraint is in the form of ``\emph{p$_{1}$ :- p$_{2}$}'' where \emph{p$_{1}$} is a component-specialization pattern and \emph{p$_{2}$} is a component pattern. For example, the following constraint should be inserted into the above constraints as below:

\begin{small}\begin{verbatim}
meaning #School -> { ...
    Teaching[*] -> {...} with { ...
        $x:#Teaching -> ($y0<=$y:Teacher, $z0<=$z:Student) :-
        $x0:.Teaching -> ($y0:Teacher, $z0:Student);... } }...
\end{verbatim}\end{small}

A complement constraint is essentially a schema mapping which can be denoted as the formula $p_{c}(\overline{x},\overline{y}) \rightarrow \exists\overline{z}.p_{cs}(\overline{x},\overline{z})$. It means that in a meaning graph for any node tuple denoted by the variable tuple $(\overline{x},\overline{y})$ that can satisfy the component relations indicated by the component pattern \emph{p$_{c}$}, there should be the certain node tuple denoted by the variable $\overline{z}$ that the node tuple denoted by $(\overline{x},\overline{z})$ can satisfy the component and the specialization relations indicated by the component-specialization pattern \emph{p$_{cs}$}.
As the above constraint shows, for each pair of the nodes \emph{(\$y0,\$z0)} of the ``\emph{Teacher}'' and ``\emph{Student}'' components of the ``\emph{\#School.Teaching}'' meaning, there is a ``\emph{\#Teaching}'' node \emph{\$x} whose ``\emph{Teacher}'' and ``\emph{Student}'' nodes indicated by \emph{\$y} and \emph{\$z} are the super-nodes of \emph{\$y0} and \emph{\$z0} respectively. Therefore, the ``\emph{\#Teaching}'' nodes implicitly introduced by the component specialization in ``\emph{\#School}'' can be instantiated with the nodes of  ``\emph{\#School.Teacher}'' and ``\emph{\#School.Student}'', which prevent the meaning graph from the incompleteness problem.

Generally, for a meaning \emph{l}, the specialization declaration and the complement constraints it is involved in are gathered as its specialization specification. A specialization specification is \textbf{\emph{complete}} iff: for any meaning graph which satisfies the specialization declarations and the complement constraints, there is not the incompleteness problem incurred by the component specialization.
SemGraph has the following property to check the validity of the specialization specification in designing the concrete models.
\begin{proposition}
Whether a specialization specification is complete is decidable in PTIME.
\end{proposition}

As the example shows, the specialization specification are listed as the constraint rules in the \emph{with} clause. SemGraph adopts the specialization constraints to make the specialization be dynamically specified. The two propositions show that the dynamically specialization is feasible since it can always check whether the new specialization relation is valid.

SemGraph also supports the recursive defined meanings by allowing the components to specialize the composite meaning. For example, a natural number meaning can be defined as follows:
\begin{small}\begin{verbatim}
intensive meaning #Nat ->
    { static Zero | public Pred with {Pred !<= #Nat} }
\end{verbatim}\end{small}
In this example, an intensive ``\emph{\#Nat}'' meaning can have a static component ``\emph{Zero}'' or a public component ``\emph{Pred}''. The ``\emph{Zero}'' component node is shared by all the ``\emph{\#Nat}'' nodes containing it. The ``\emph{Pred}'' component would have a ``\emph{Zero}'' component or another ``\emph{Pred}'' component, since it is a sub meaning of ``\emph{\#Nat}''. As the ``\emph{\#Nat}'' are declared to be intensive,  two ``\emph{\#Nat}'' nodes are equivalent to each other if they contain the equivalent components. Such a recursive definition using the specialization can exactly generate a set of nodes corresponding to the natural number type, i.e., ``\emph{0:\#Nat$\rightarrow$Zero}'', ``\emph{1:\#Nat$\rightarrow$Pred$\rightarrow$Zero}'', ``\emph{2:\#Nat$\rightarrow$Pred$\rightarrow$Pred$\rightarrow$Zero}'', etc.

%
%


The reference relation indicates an association between two independent meanings. In a reference relation like ``\emph{r:\#School.President $\dashrightarrow$ e:\#School.Teacher}'',  the node \emph{r} of the ``\emph{\#School.President}'' is associated with, or in other words, \textbf{\emph{refers to}}, the node \emph{e} of the ``\emph{\#School.Teacher}''. However, this association is different from the one of specialization in that it is alterable. That is, during the semantics evolution, the node \emph{r} can refer to another node \emph{s} of the ``\emph{\#School.Teacher}''.
Here the label ``\emph{\#School.President}'' is called a \emph{reference meaning} and the label ``\emph{\#School.Teacher}'' is called the \emph{host meaning} of ``\emph{\#School.President}''.

The reference meanings are analogous to the reference types comprehensively used in common programming languages to allocate the space for the mutable values. In fact,  SemGraph implicitly introduces a ``pure'' reference meaning denoted as ``\emph{@a}'' for each non-reference meaning \emph{a}, and the reference meaning declared by the users are the sub meaning of these ``pure'' meanings. Additionally, the reference meaning \emph{@a} can automatically have a super meaning \emph{@b} if \emph{a} is the sub meaning of \emph{b}. Therefore, it is safe to use a meaning node referred to by the reference node of \emph{@a} as a meaning node of \emph{b}.

For example, in the School meaning schema a reference meaning should be declared as ``\emph{.President$\dashrightarrow$.Teacher}''.
This declaration claim the ``\emph{\#School.President}'' be a sub meaning of ``\emph{@\#School.Teacher}''. Meanwhile, since the declaration all use the abbreviated meaning name, it indicates that only the ``\emph{Teacher}'' node in the current ``\emph{\#School}'' node can be the host of the ``\emph{\#School.President}''.

\section{Semantic Data Modeling with SemGraph}
\subsection{Semantic type construction}

Constructing the compound semantic types to present complex meanings is the most important task for a semantic data model. Existing data models usually provide some common constructors like the product (or aggregation) and the group, meanwhile some models also try to use the preliminary generalization mechanism to implement the union type.

SemGraph can easily implement various type constructors, since the composition relation and the component occurrence mechanisms has already laid the foundation of the product(or aggregation), the group and the union types. SemGraph especially provides a product type constructor ``\^{}'', a group type constructor ``*'' and a union type constructor ``$||$'', to explicitly declare a compound meaning type of other root meanings. (A compound meaning is a special composite meaning generated by the constructors.)
For example, the compound meaning ``\emph{\#School\^{}\#GIS}'' consists of two components ``\emph{\#School\^{}\#GIS.School}'' and ``\emph{\#School\^{}\#GIS.GIS}'' which are respectively the sub meanings of the two root meanings ``\emph{\#School}'' and  ``\emph{\#GIS}'', and the two components should occur conjunctively. For the compound meaning ``\emph{*\#School}'', the component meaning ``\emph{*\#School.School}'' should be declared to have multiple instances. For the compound meaning ``\emph{\#School$||$\#Institute}'', it is the super meaning of  ``\#School'' and ``\#Institute'' whose components should occur disjunctively in the union.

For the above constructors, SemGraph does not support the compound meaning of the components of other root meanings, because the existence the components  always assume the existence of their root meanings and thus the compound of their root meanings should be declared previously.


Although the compound meanings can easily implement the functions of the conventional type constructors, they are different in their intension. The latter like the aggregation often represents the loose associations of the member entities or relations, and the member types don't depend on the compound semantic types.
However, the former is declared as the context under which the components can make sense. Therefore we restrict the components of the product or group meaning to be the sub meanings of the original root meanings.

Intersection of semantic concepts is usually considered by the logical based semantic models, e.g., the ontology models, and seldom directly supported by the data models. Some data models can indirectly implement the limited features of the intersection through multiple inheritance.
SemGraph introduces a special intersection compound meaning using an intersection constructor ``\&''. For a intersection meaning \emph{l$_{a}$\&l$_{b}$}, the meanings \emph{l$_{a}$} and \emph{l$_{b}$} are both the sub meanings of a certain meaning \emph{l}, and \emph{l$_{a}$\&l$_{b}$} is automatically declared as the sub meaning of \emph{l$_{a}$} and \emph{l$_{b}$} respectively. The  \emph{l$_{a}$\&l$_{b}$} meaning inherits the components of its super meanings (through a certain mechanism avoiding the name conflict). Recursively, \emph{l$_{a}$\&l$_{b}$} would automatically declare the certain intersection component meanings  \emph{l$_{1}$\&l$_{2}$} where the meaning \emph{l$_{1}$} and \emph{l$_{2}$} are respectively the components of \emph{l$_{a}$} and \emph{l$_{b}$} and they are both the sub meanings of a component \emph{l'} of \emph{l}. At the instance level, a node of \emph{l$_{1}$\&l$_{2}$ } certainly shares the same super-node of \emph{l'} as \emph{l$_{1}$} and \emph{l$_{2}$}.

For example, for the meanings ``\emph{\#School}'' and ``\emph{\#Institute}'' which are both the sub meanings of ``\emph{\#Organization}'', the component meaning ``\emph{\#Organization.Member}'' is specialized as the meanings ``\emph{\#School.Teacher}'' and ``\emph{\#Institute.Researcher}'', and thus the intersection ``\emph{\#School\&\#Institute}'' will automatically declare a new component ``\emph{Teacher\&Researcher}''. In the meaning graph, if a node of ``\emph{\#Organization}'' has a sub-node of ``\emph{\#School}'' and a sub-node of ``\emph{\#Institute}'', they would have a
sub-node of ``\emph{\#School\&\#Institute}'' which not only inherits nodes of the components ``\emph{Teacher}'' and ``\emph{Researcher}'' but also has the new nodes of ``\emph{Teacher\&Researcher}'' which are the sub-nodes of the ``\emph{Teacher}'' nodes and ``\emph{Researcher}'' nodes sharing the common ``\emph{Member}'' super-nodes.

To avoid the possible combinatorial explosion of the compound meaning types produced by the type constructors, SemGraph deploys a declare-to-use policy. That is, the compound meanings with the type constructors only make sense if they are explicitly declared in the schema. Furthermore,  since the intersection meaning only works as the two super meanings \emph{l$_{a}$} and \emph{l$_{b}$} have the common ancestor \emph{l}, the amount of the intersection node and its component nodes can be restricted in the polynomial space.


\subsection{Attribute and role}

The attribute is widely adopted in the semantic data models and the object-oriented data models. An attribute is actually a function carrying the certain semantic information of the source object. An attribute can be either single valued or multi-valued.
In SemGraph, an attribute is easily represented as a sub meaning or reference meaning being the component of the meaning representing the source object. For an attribute whose target object would not be altered, e.g., the name string of a person, it can be specified as the sub meaning like  ``\emph{t:\#Person.Name$\Leftarrow$`John':\#String}'' in Fig.1; for one with mutable target object, e.g., the age of a person, it is specified as the reference meaning like ``\emph{w:\#Person.Age$\dashrightarrow$20:\#Nat}''. Furthermore, a multi-valued attribute can be represented as a set of components or a composite component. 


Another important feature of the reference meaning is that it can be directly used to implement the role models.
Since a reference  node refers to mutable host nodes, the reference meaning can be treated as a role of the host meaning. Additionally, the component meanings of the reference meaning are the extended semantic information specific for the roles, since they only make sense under the context of the role and thus are independent to the host. For example, the above definition of the ``\emph{\#School.President}'' can be extended with ``\emph{President$\rightarrow$\{Office, Telephone\}}''. It means that the president role played by a teacher have the components ``\emph{Office}'' and ``\emph{Telephone}''  which only pertain to the role rather than the teacher who plays the role.

\subsection{Derived meaning, join and natural Join}

The derived type is a special subtype whose existence depends on the satisfaction of certain conditions on the super instances. It is supported by some semantic data models, but not by most O-O models where the sub object should coexist with its super object. SemGraph supports the derived type by specifying it as a sub meaning with a user-defined constraint, and the existence of a derived node is independent to the super-node.
Further, using the specialization and the product type constructor, SemGraph can implement a join operation of the meanings which is not often supported by the other semantic data models.

In practical semantic information gathering and analysis, an important and interesting issue is to associate the related objects to form a broader view of the related things. For example, for the two meaning ``\emph{\#School}'' and ``\emph{\#GIS}'', we might be interested in the combination of the two meanings which is linked by the locations in the school. This issue is well studied as the natural join operation in conventional databases, however, it is not thoroughly solved in the semantic data modeling. A possible solution is to treat it as common join operations, however, the natural join has different semantic features from common join since its linking joints should semantically be the same thing and the join result is not so volatile as the derived types.

In SemGraph the natural join operation is presented as a specialization of the product compound type with the new component indicating the linking joint.
For the two meanings \emph{l} and \emph{l'} where there are the component meanings  \emph{l.x} and \emph{l'.y} and \emph{x} and \emph{y} share a common ancestor meaning \emph{z}, the natural join of \emph{l} and \emph{l'} through \emph{l.x} and \emph{l'.y} is presented as the sub meaning of the product meaning \emph{l\^{}l'} where there exists a component \emph{l.x\&l'.y}, and thus is denoted as \emph{l\^{}l'$_{l.x\&l'.y}$}. Since \emph{l.x\&l'.y}  shares the super meaning \emph{z} with \emph{l.x} and \emph{l'.y}, the join meaning \emph{l\^{}l'$_{l.x\&l'.y}$} can exactly combine \emph{l} and \emph{l'} using \emph{l.x\&l'.y} as the joint. For example,  the natural join of the meanings ``\emph{\#School}'' and ``\emph{\#GIS}'' can have a certain joint meaning ``\emph{\#School.location.geo\_id\&\#GIS.loc\_geo.geo\_id}'', assuming that the ``\emph{\#GIS}'' meaning be defined following the above guidelines.

\subsection{Dynamical specialization}

SemGraph supports the dynamical specialization of the meanings. Since the specialization specifications are in the form of constraints, they can be dynamically generated between two meanings, only if the specification is complete and can be satisfied. SemGraph uses a ``specialize'' statement to explicitly declare a dynamical specialization, which can be used to implement the generalization of the existing schema.
\begin{small}\begin{verbatim}
meaning #Organization -> { Member[*], Head } with { .Head --> .Member}
specialize #School <=! #Organization -> {
      .Teacher <=! .Member; .Student <=! .Member; .President <=! .Head}
\end{verbatim}\end{small}

As the above example shows, an ``\emph{\#Organization}'' meaning is defined after the ``\emph{\#School}'' meaning. It is dynamically declared to be the super meaning of ``\emph{\#School}'' and the ``specialize'' statement uses the nested declarations to specify the specialization relations between the components. Since it is essentially a generalization of the existing schema, the ``$\Leftarrow$!'' specialization is used here to show that the super-node should coexist with the sub-node since its existence depends on the sub-node's.

\section{Conclusion}

Semantic information comes from the subjective views of the things and is usually recognized in a posterori way, but it is often represented by the semantic models under a priori assumption following an entity-relationship-oriented approach. 
In this paper, we have proposed a meaning-oriented approach to semantic data modeling and proposed a 
new graph-based semantic data model named SemGraph following this approach.
In comparison with the previous studies, SemGraph deploys the simple but expressive mechanisms of the composition, equivalence, specialization and reference to carry out both common and specific requests on semantic data modeling.

Our work on the meaning-oriented approach is in the preliminary stage. Based on the definitions and the mechanisms listed in the paper, we are studying the issues of enhancing the models to specify the semantics of regular structure patterns and the spatial-temporal information. Furthermore, we would develop a full-fledge query language that can explore the meaning graphs and extract meaningful results.


\begin{thebibliography}{20}
\bibitem{survey1}
R.Hull, R.King. Semantic database modeling: Survey, applications, and research issues. ACM Comput. Surv. 19(3):201-260, 1987.
\bibitem{survey2}
J.Peckham, F.J. Maryanski. Semantic data models. ACM Comput. Surv. 20(3), 153-189, 1988.
\bibitem{ifo}
S. Abiteboul,  R. Hull. IFO: A formal semantic database model. ACM Transactions on Database Systems (TODS) 12(4):525-565, 1987.
\bibitem{sdm}
M.Hammer, D.Mcleod. Database description with SDM: A semantic database model. ACM Trans. Database Syst. 6(3):351-386, 1981.

\bibitem{3}
M.P.Atkinson, F.Bancilhon, D.J.De Witt, et-al. The object-oriented database system manifesto.  Proc. of SIGMOD, 1989.
\bibitem{4}
A.A.G.Ghelli, R.Orsini. A relationship mechanism for a strongly typed objectoriented database programming language. Proc. of VLDB'91,  565-575, 1991.


\bibitem{16}
R.G.Cattell, D.Barry, M.Berler, J.Eastman, et-al. The Object Data Standard: ODMG 3.0. Morgan
Kaufmann Publishers, San Francisco, 2000.
\bibitem{13}
F.Steimann. On the representation of roles in object-oriented and conceptual
modelling. Data Knowledge Engineering 35(1):83-106, 2000.
\bibitem{inm}
M.Liu, J.Hu. Information Networking Model. Proc. of ER'09, 131-144, 2009.
\bibitem{gsurvey1}
R.Angles, C.Gutierrez. Survey of graph database models. ACM Computing Surveys, 40(1):1, 2008.
\bibitem{gdm}
J.Hidders.  Typing Graph-Manipulation Operations. Proc. of ICDT'02, 394-409, 2002.
\bibitem{groovy}
M.Levene,  G.Loizou. A Graph-Based Data Model and its Ramifications. IEEE
Transactions on Knowledge and Data Engineering, 7(5):809-823, 1995.
%



\bibitem{gdm1}
C. Sankhayan, N. Chaki, et-al. Gdm: a new graph based data model using functional abstractionx. Journal of Computer Science and Technology 21(3):430-438, 2006.

\bibitem{rdfs}
D.Brickley, R.Guha. RDF Schema 1.1. http://www.w3.org/TR/rdf-schema/, 2014.

\bibitem{lod}
T. Heath, C. Bizer. Linked Data: Evolving the Web into a Global Data Space. Synthesis Lectures on the Semantic Web, 1:1, 1-136. Morgan \& Claypool, 2011.

\bibitem{graphdb}
R. H. Guting.  GraphDB: Modeling and Querying Graphs in Databases. VLDB'94, 297-308. 1994.



\bibitem{gsurvey2}
R. Angles. A comparison of current graph database models. In Data Engineering Workshops (ICDEW), pp. 171-177. 2012.

\bibitem{graphdata}
Buerli M. The Current State of Graph Databases. Department of Computer Science, Cal Poly San Luis Obispo, mbuerli@ calpoly. edu, 2012.




\bibitem{ghafoor}
W. M. Ahmed, D. Lenz, J. Liu, et-al. XML-Based Data Model and Architecture for a Knowledge-Based Grid-Enabled Problem-Solving Environment for High-Throughput Biological Imaging. IEEE Transactions on Information Technology in Biomedicine, 12(2):226-240, 2008.

\bibitem{xmlsem}
M.Murali, D.Lee,  R.R.Muntz. Semantic data modeling using XML schemas. Proc. of ER'01, 149-163, 2001.


\end{thebibliography}
\end{document}